\documentclass[11pt]{article}
\usepackage{multicol}
\usepackage{graphicx}
\usepackage{amssymb}
\usepackage{epsfig}

\def\bea{\begin{eqnarray}}
\def\eea{\end{eqnarray}}
\begin{document}
\title{\bf Quasi-Rip universe induced by the fluid with inhomogeneous equation of state}
\author{I. Brevik${}^1$, V.V Obukhov${}^2$, A.V. Timoshkin${}^3$\\
\\
${}^1$ Department of Energy and Process Engineering,\\
Norwegian University of Science and Technology, N-7491 Trondheim, Norway.\\
${}^{2,3}$ Department of Theoretical Physics,\\
Tomsk State Pedagogical University, Tomsk, 634061, Russia.\\
\\
iver.h.brevik@ntnu.no obukhov@tspu.edu.ru timoshkinAV@tspu.edu.ru}
\date{}

\maketitle

\begin{abstract}
    We investigate a specific model for dark energy, which lead to the Quasi-Rip cosmology.
    In the Quasi-Rip model, the equation of the state parameter $w$ is less than $-1$
    in the first stage, but then in the second stage is larger than $-1$.
    The conditions for the appearance the Quasi-Rip in the terms
    of the parameters equation of state are received.
\end{abstract}

%% \MSC 78A35 \sep 78-05 \sep 81V45

\begin{multicols}{2}
%\vspace{-10mm}
\section{Introduction}

The discovery of an accelerated expansion of the universe led to
the appearance the new theoretical models in the cosmology (for
recent review see [1,2]) and significantly changed our view of the
fate of the universe. Recent observations suggest that the
universe is dominated by a negative-pressure component named dark
energy (dark fluid). Such quintessence/phantom dark energy
proposed to explain the cosmic acceleration should have the strong
negative pressure. It can be characterized by an equation of state
parameter $w$, which is the ratio of the pressure to the density:
$w=\frac{p}{\rho}$. Its equation of state  parameter $w$ is
smaller than $-1$. The condition $w<-1$ corresponds to a dark
energy density that monotonically increases with time $t$ and
scale factor $a$. There are some interesting possible scenarios
concerning the fate of the universe, including Big Rip [3-4],
Little Rip [5-12], Pseudo-Rip [13] models. These models are based
on the assumption that the dark energy density $\rho$ is a
monotonically increasing function. In this paper we are interested
in studying a cosmological model in which the dark energy density
$\rho$ monotonically increases $\left( w<-1\right)$  in the first
stage and then monotonically decreases $\left( w>-1\right)$. At
the first stage, there takes place a disintegration of bound
structures, but then at the second stage the disintegration
process will stop, and the disintegrated structures have the
possibility to be recombined. This scenario in cosmology was
called the Quasi-Rip [14]. Note that models for which $\rho$ is
not monotonic are physically less plausible, and it is more
difficult to make any general statement about such models. The aim
of this article is to examine the influence of dark fluid equation
of state explicitly dependent on scale factor $w$ and $\Lambda$ on
the occurrence of the Quasi-Rip.

\section{Dark fluid with inhomogeneous equation of state in the Quasi-Rip model}\label{sec2}
%
%%%%%%%%%%%%%%%%%%%%%%%%%%%%%%%%%%%%%%
Let us consider an explicit model of the Quasi-Rip. We choose to the energy density  $\rho$ as a function of the scale factor $a$. The simplest function  $\rho$ can be given [14]:
\begin{equation} \label{rho}
\rho=\rho_0 a^{\alpha-\beta\ln a},
\end{equation}
where $\rho=\rho_0$ at fixed time $t_0$, $\alpha$ and $\beta$ are both constants.
%% ссылка на эту формулу так: формула (\ref{rho})
The derivative of the energy density with respect to cosmic time $t$ is equal:
\begin{equation} \label{a}
\rho'=\rho H \left(\alpha-2\beta\ln a\right)
\end{equation}
where $H=\frac{a'}{a}$  is the Hubble rate.
We assume, that our universe is filled with an ideal fluid (dark energy) obeying an inhomogeneous equation of state [15]:
\begin{equation} \label{b}
\ p=\ w \left( a\right) \rho +\Lambda \left(a\right)
\end{equation}
where $ w \left( a\right)$ and $ \Lambda \left(a\right)$ depend on the scale factor $a$,  $p$ is the pressure.
Let us write down the energy conservation law:
\begin{equation} \label{c}
\rho'+3 H \left( p+\rho\right) =0
\end{equation}
Taking into account (2-4) we obtain:
\begin{equation} \label{d}
\rho \left(\alpha-2\beta\ln a\right) +\rho\left( 1+w \left( a\right)\right) +\Lambda \left(a\right)=0
\end{equation}
Let us express from the equation (5) $w \left( a\right)$ and suppose, that the cosmological constant $\Lambda$  is proportional to the square of the energy density, that is:
\begin{equation} \label{e}
\Lambda \left(a\right) = \gamma\rho^{2}_{0} a^{2\left( \alpha-\beta\ln a\right) },
\end{equation}
where  $\gamma$ is a some constant.
Using (6), one obtains:
\begin{equation} \label{f}
w \left( a\right)= -1 -\gamma\rho_{0} a^{ \alpha-\beta\ln a} +2\beta \ln a -\alpha
\end{equation}
If we require $\beta>0$, then the extremum of $\rho$ is a maximum. It reaches at $ a=\exp {\frac{\alpha}{2\beta}} $, then the parameter is equal:
\begin{equation} \label{g}
w \left( a\right)= -1 -\gamma\rho_0^\frac{\alpha}{2}
\end{equation}
Consequently, if we take an ideal fluid, obeying an equation of state (3) and (5), then obtain the solution, which realize the Quasi-Rip (1). Note, that such Quasi-Rip is caused exceptionally with cosmological constant $\Lambda$.
Let us solve the equation (5) about  $\Lambda$ and choose the parameter  $w \left( a\right)$ in the view:
\begin{equation} \label{h}
w \left( a\right)= -1 - \frac{\delta}{3\rho_0} a^{\beta\ln a-\alpha},
\end{equation}
where $\delta$ is a constant.
We obtain the expression:
\begin{equation} \label{k}
\Lambda \left(a\right)= -\rho_0 a^{\alpha-\beta\ln a} \left( \alpha-2\beta\ln a\right) -\frac{\delta}{3}
\end{equation}
In this case the Quasi-Rip is caused in the part $w$. The future behavior of our universe can depending on the particular model parameters $\alpha$ and $\beta$.
Thus, we explored the equation of state (3), which gives the Quasi-Rip.
 %%%%%%%%%%%%%%%%%%%%%%%%%%%%%%%%
\section{Conclusion}
%%%%%%%%%%%%%%%%%%%%%%%%%%%%%%%%
The Quasi-Rip model has an unique feature different from Big Rip, Little Rip and Pseudo-Rip. All these models arise from the assumption, that dark energy density  is monotonically increasing. They lead to the dissolution of all bound structures. As distinct from these models in Quasi-Rip model this assumption is broken. Our universe has a possibility to be rebuilt after the rip.  In present work we have built the Quasi-Rip universe induced by the dark fluid with inhomogeneous equation of state. It is shown, that the Quasi-Rip cosmology can be caused exceptionally with the cosmological constant or the parameter . It would be interesting to understand if Quasi-Rip cosmology may be mapped with dark energy fluid cosmology mimicking string-landscape features [16]. From another side, the role of viscosity in Rip cosmology may be also relevant in Quasi-Rip picture [17].
%%%%%%%%%%%%%%%%%%%%%%%%%%%%%%%%
\section*{Acknowledgement}
We are grateful to professor Sergei Odintsov for helpful discussions.

%%%%%%%%%%%%%%%%%%%%%%%%%%%%%%%%

\end{multicols}

\begin{thebibliography}{99}

\bibitem{Robin} K. Bamba, S. Capozzielo, S. Nojiri, S.D. Odintsov. arXiv: 1205.3421v3[gr-qc].

\bibitem{Drob}S. Nojiri, S.D. Odintsov. Phys. Rept.505: 59-144, 2011, arXiv:1011.0544v4[gr-qc].


[arXiv:0710.0288 [astro-ph]].
\bibitem {Lindhard} R.R. Caldwell, M. Kamionkowski, N.N. Weinberg, Phys. Rev. Lett. 91, 071301, 2003 [astro-ph/0302505];
\bibitem {No} S. Nojiri, S.D. Odintsov, Phys. Rev. D70, 103522, 2004; Phys. Lett. B562, 147, 2003.
\bibitem {Baz} H. Frampton, K.J. Ludwick, R.J. Scherrer. Phys. Rev.D84: 063003, 2011, arXiv:1106.4996v1[astro-ph.CO].

\bibitem {Bre} I. Brevik, E. Elizalde, S. Nojiri, S.D. Odintsov. arXiv: 1107.4642v2[hep-th].

\bibitem {Fla} P.H. Frampton, K.J. Ludwick, S. Nojiri, S.D. Odintsov, R.J. Scherrer. Phys. Lett. B708, (2012), 204-211, arXiv: 1108.0067v2[hep-th].
\bibitem {Ast} A.V. Astashenok, S. Nojiri, S.D. Odintsov, A.V. Yurov. arXiv: 1201.4056v2[gr-qc].
\bibitem {No} A.V. Astashenok, S. Nojiri, S.D. Odintsov, R.J. Scherrer. arXiv: 1203.1976v2[gr-qc].
\bibitem { El} A.V. Astashenok, E. Elizalde, S.D. Odintsov, A.V. Yurov. arXiv: 1206.2192v1[gr-qc].
\bibitem {Od}  S. Nojiri, S.D. Odintsov, D. Saez-Gomez. arXiv: 1108.0767v2[hep-th].
\bibitem  {Mak} A.N. Makarenko, V.V. Obukhov, I.V. Kirnos. arXiv: 1201.4772v2[gr-qc].
\bibitem {Kil} P.H. Frampton, K.J. Ludwick, R.J. Scherrer. Phys. Rev. D85: 083001, 2012, arXiv:1112.2964v2[astro-ph. CO].
\bibitem{Berna}  H. Wei, L.F. Wang, X.J. Guo. arXiv: 1207.2898.v1[gr-qc].
\bibitem {Landau}  S. Nojiri, S.D. Odintsov. Phys. Rev.D72, 023003, 2005, arXiv: 0505215[hep-th].
\bibitem {Bor} E. Elizalde, A.N. Makarenko, S. Nojiri, V.V. Obukhov, S.D. Odintsov. arXiv: 1206.2702[gr-qc].
 \bibitem {Ob} I. Brevik, R. Myrzakulov, S. Nojiri, S.D. Odintsov. Phys. Rev.D86: 063007, 2012, arXiv: 1208.4770[gr-qc].
\end{thebibliography}
\end{document}